\documentclass[twocolumn]{aastex701}

\usepackage{caption}
\usepackage{subcaption}

\begin{document}

\title{Surface regolith model of near-Earth asteroid (3200) Phaethon}

\author[orcid=0009-0007-9187-6460, gname=Mikko, sname=Vuori]{Mikko Vuori}
\affiliation{Department of Physics, P.O. box 64, FI-00014 University of Helsinki, Finland}
\email[show]{mikko.vuori@helsinki.fi}

\author[orcid=0000-0001-7403-1721, gname='Antti', sname='Penttilä']{Antti Penttilä}
\affiliation{Department of Physics, P.O. box 64, FI-00014 University of Helsinki, Finland}
\email{antti.i.penttila@helsinki.fi}

\author[orcid=0000-0001-8058-2642, gname='Karri', sname='Muinonen']{Karri Muinonen}
\affiliation{Department of Physics, P.O. box 64, FI-00014 University of Helsinki, Finland}
\email{karri.muinonen@helsinki.fi}

\author[orcid=0009-0001-9054-5068, gname='Hanna', sname='Pentikäinen']{Hanna Pentikäinen}
\affiliation{Department of Physics, P.O. box 64, FI-00014 University of Helsinki, Finland}
\email{hanna.pentikainen@helsinki.fi}

\author[orcid=0009-0008-7221-1671, gname='Ari', sname='Leppälä']{Ari Leppälä}
\affiliation{Department of Physics, P.O. box 64, FI-00014 University of Helsinki, Finland}
\email{ari.leppala@helsinki.fi}

\author[orcid=0000-0002-6610-1897, gname=Zuri, sname=Gray]{Zuri Gray}
\affiliation{Department of Physics, P.O. box 64, FI-00014 University of Helsinki, Finland}
\email{zuri.gray@helsinki.fi}

\author[orcid=0000-0002-9870-123X, gname=Eric, sname=MacLennan]{Eric MacLennan}
\affiliation{Department of Physics, P.O. box 64, FI-00014 University of Helsinki, Finland}
\affiliation{Instituto de Astrofísica de Canarias (IAC), 38205 La Laguna, Tenerife, Spain}
\affiliation{Departamento de Astrofísica, Universidad de La Laguna 38205 La Laguna, Tenerife, Spain}
\email{eric.maclennan@iac.es}

\author[orcid=0000-0002-5624-1888, gname=Mikael,sname=Granvik]{Mikael Granvik}
\affiliation{Department of Physics, P.O. box 64, FI-00014 University of Helsinki, Finland}
\affiliation{Asteroid Engineering Laboratory, Luleå University of Technology, Box 848, SE-98128 Kiruna, Sweden}
\email{mikael.granvik@helsinki.fi}

\author[orcid=0000-0002-4129-5381, gname='Anne', sname='Virkki']{Anne Virkki}
\affiliation{Department of Physics, P.O. box 64, FI-00014 University of Helsinki, Finland}
\email{anne.virkki@helsinki.fi}

\begin{abstract}

    The near-Earth asteroid (3200) Phaethon displays activity as it comes close to the Sun and is presumed to be the progenitor of the Geminids meteor shower.
    Both CM- and CY-meteorites have been suggested as possible meteorite analogues for Phaethon.
    Previous spectral and polarization measurements of Phaethon indicate possible surface heterogeneity in the regolith properties.
    By combining mineral components and their size fractions, we present a regolith model for Phaethon that reproduces the spectral, polarization, and geometric albedo measurements presented in various studies, with the aim of quantitatively characterizing the surface material and structure, as well as the regolith heterogeneity.
    The best fit model able to produce the measurements is constructed from large olivine particles with troilite inclusions and small scale surface roughness, mixed with smaller, porous troilite aggregates comprised of monomers resembling interplanetary dust. Our model suggests that the different spectral and polarization responses of the regolith are caused by changes in the size and amount of the small dark troilite particles. Based on the amount of troilite in the regolith model, we propose that CM-meteorites are the preferable meteorite analogue for Phaethon.

\end{abstract}

\section{Introduction} \label{sec:introduction}

    (3200) Phaethon is an unusual potentially-hazardous near-Earth asteroid \citep{LAZZARIN2019}, and a target of the JAXA DESTINY$^+$ space mission \citep{OZAKI2022}.
    Phaethon's perihelion distance is $q=0.14\,$au, where it is exposed to extreme heating \citep{GIBSON1983, HANUS2016,MacLennan_etal21}.
    Phaethon shows activity as brightening near its perihelion passage \citep{JEWITT2010} caused by dust \citep{JEWITT2013, LI2013} or gas \citep{HUI2023, ZHANG2023} emission, or a combination of both \citep{MACLENNAN2024}.
    Phaethon has also been suggested as the progenitor of the Geminid meteor shower, as its orbit coincides with the Geminids \citep{WHIPPLE1983, CUKIER2023}. Asteroid 2005 UD has been proposed to be a fragment of Phaethon as it follows a similar orbit and shows a similar polarization response and composition to Phaethon \citep{OHTSUKA2006, ISHIGURO2022, DEVOGELE2020}, although the dynamical connection remains uncertain \citep{RYABOVA2019}.
    Phaethon bears resemblance to (2) Pallas and its collisional family with a similar blue spectrum classified as belonging to a carbonaceous B-type asteroid \citep{LICANDRO2007, DELEON2010}.
    Phaethon's surface has been proposed to be composed of material that is analogous to either CI/CM or CY meteorites \citep{King_etal21t,KING2019,SUTTLE2023}, meaning some of these meteorites could originate from Phaethon \citep{LICANDRO2007, LAZZARIN2019, SUTTLE2024, THOMAS2025, ANGRISANI2025, MACLENNAN2024}.
    
    Differences in spectral and polarization measurements suggest that regolith grain properties across Phaethon's surface are heterogeneous.
    Spectral observations have suggested both rotational \citep{OHTSUKA2020} and latitudinal heterogeneity \citep{KARETA2018,LAZZARIN2019,LEE2019,BETZLER2024}, with the latter also supported by an analysis accounting for photometric correction \citep{ANGRISANI2025}.
    Radar, thermal, and geophysical studies also support surface heterogeneity.
    \cite{TAYLOR2019} identified a possible crater and a radar-dark region, \cite{MACLENNAN2022} reported differences in thermal inertia between the hemispheres, and \cite{JO2025} suggested freshly exposed material in the northern hemisphere.
    Polarimetric observations have also suggested rotational heterogeneity \citep{BORISOV2018}.
    The most alluring evidence for heterogeneity is the extreme maximum of the degree of linear polarization reported in \cite{ITO2018}, which has not been found in any other polarization observation \citep{DEVOGELE2018,SHINNAKA2018}.
    
    In \citet{GRAY2026}, new polarimetric observations of Phaethon are presented.
    The authors study whether the differences in the polarization observations can be explained with the surface regions sampled during different apparitions.
    They found that the unusual high polarization of \citet{ITO2018} originates from the northern hemisphere,
    and their new observations similarly sample mostly the northern hemisphere.
    However, their new observations showed an intermediate polarization response between the previous polarization responses.
    They proposed that observation geometry can play a role, as the unusual high polarization observation samples the asteroid surface at systematically different incidence and emergence angles.
    The authors also suspect that the more extreme observation geometry compared to \citet{ITO2018} has a depolarizing effect.
    In general, the observations and analysis in \citet{GRAY2026} suggest hemispherical regolith heterogeneity.
    
    The surface regolith of an asteroid can be studied from its photometric, spectral, and polarimetric response.
    Classically, an asteroid's spectrum provides information on the surface composition, and linear polarization on the grain size.
    The linear polarization phase curve of an asteroid has features it can be quantified by: the minimum polarization degree in the backscattering region ($P_{\text{min}}$) and its phase angle ($\alpha_{\text{min}}$), the inversion angle where linear polarization changes from negative to positive ($\alpha_0$) and the slope of the curve at that angle ($h$), and the degree of maximum polarization ($P_{\text{max}}$) and its phase angle ($\alpha_{\text{max}}$). Observations can be analyzed, for example, by looking at features such as absorption bands in a spectrum \citep{REDDY2015, DEMEO2022}, linking multiple scattering of small particles to the negative branch of the linear polarization \citep{SHKURATOV2002, BACH2024}, and the derivation of the geometric albedo from the polarization phase curve \citep{CELLINO2015}. The light scattering effects that cause such features in the spectrum and linear polarization are unfortunately not as simple and straightforward (e.g., \citet{SHKURATOV2005, HAPKE2008, KIDDEL2018, FRATTIN2019, SPADACCIA2022, FRATTIN2025}), and is especially true for Phaethon which has a completely featureless blue spectrum and an anomalous polarization curve. To understand the regolith that causes the spectral and polarimetric response requires inverse light scattering modeling \citep{MUINONEN2019, PENTTILA2024}.
    
    Light scattering modeling of Phaethon's polarization phase curve has previously been reported by \cite{KISELEV2022}. Using a mixture of $\sim1$~{\textmu}m sized Gaussian particles of forsterite and amorphous carbon, their model accurately reproduces the observed polarization curve at phase angles $\alpha=20^{\circ}$--$130^{\circ}$.
    Spectral modeling reported by \cite{LICANDRO2007} reproduced the spectrum and albedo of Phaethon in the Vis-NIR spectral range using a mixture of two hydrated minerals, amorphous carbon, and graphite.
    
    In this article, we present light scattering modeling of Phaethon's surface regolith based on a combination of spectral, linear polarization, and geometric albedo measurements.  In addition, we use our model to investigate possible interpretations of the observed differences in spectral and polarization measurements reported by various groups, and explore their implications for surface heterogeneity and meteorite analogues.

\section{Methods}

    \subsection{Observational constraints} \label{sec:obs-data}
        We use a subset of the available observational data to compare and constrain the light scattering model.
        In our analysis, we include spectral observations from \cite{LAZZARIN2019} and the MITHNEOS survey \citep{BINZEL2019}\footnote{The spectrum is an unpublished 2024 spectrum from the MITHNEOS MIT-Hawaii Near-Earth Object Spectroscopic Survey.}.
        The \citet{LAZZARIN2019} spectral dataset displays the greatest variation in the UV-Vis wavelength range, while the MITHNEOS spectrum provides a broad wavelength range from UV to NIR.
        Other spectral observations have been published, for example, by \cite{KARETA2018}, \cite{LEE2019}, \cite{OHTSUKA2020}, \cite{TAKIR2020}, \cite{KISELEV2022}, and \cite{THOMAS2025}.
        These published spectra are broadly consistent with the observations of \cite{LAZZARIN2019} and do not show substantially different spectral behavior.
        We therefore focus on the \cite{LAZZARIN2019} and MITHNEOS datasets, which span the largest range of spectral properties and provide the clearest evidence for spectral heterogeneity.

        As for linear polarization, we use data from \cite{ITO2018}, \cite{SHINNAKA2018}, \cite{DEVOGELE2018}, \cite{OKAZAKI2020}, \cite{GEEM2022}, \cite{KISELEV2022}, and \cite{GRAY2026} in our analysis. 
        
        The current estimates for the geometric albedo of Phaethon in the V-band range from 0.06 to 0.16 \citep{TEDESCO2002, HANUS2016, ITO2018, KARETA2018, SHINNAKA2018, MASIERO2019, KISELEV2022, GEEM2022, GRAY2026}.
        In our light scattering model, we fit the albedo at 0.56~{\textmu}m within the V-band albedo estimation range.

    \subsection{Light scattering modeling}

        To simulate spectra and polarization from Phaethon, we utilize two light scattering simulation codes: SIRIS4 \citep{MUINONEN2009, LINDQVIST2018} and RT-CB (Radiative Transfer and Coherent Backscattering) \citep{MUINONEN2004, MUINONEN2012, MUINONEN2024, MUINONEN2025}.
        We compute the scattering properties of single particles using Mie code absMie and discrete-dipole approximation code ADDA \citep{YURKIN2011}.
        SIRIS4, RT-CB, and absMie are available as open source\footnote{The software repository of the Planetary-System Research group at the University of Helsinki at \url{https://bitbucket.org/planetarysystemresearch}}.
        
        The ADDA software package is capable of DDA calculations in a distributed computing environment using a given arbitrary shape for the particle.
        In DDA, the shape of the particle is constructed from small cubical dipoles.
        DDA is a numerically exact method, where light scattering effects are calculated from the dipoles by solving a system of linear equations \citep{YURKIN2007}.
        We use ADDA to simulate particles that are too small for ray tracing, but for which the shape information is needed, and Mie calculations using spheres is not a reasonable approximation.
        
        SIRIS4 is a geometric optics code with diffuse scatterers. It calculates scattering efficiencies and scattering matrices for a given particle shape or a Gaussian random particle \citep{MUINONEN1996, MUINONEN2009}.
        We first use SIRIS4 to calculate the scattering properties of large single particles, which are used in multi-particle simulations.
        Even though SIRIS4 calculates scattering from single particles, it is also utilized for multi-particle simulations using the method described in \cite{MARTIKAINEN2018}.
        SIRIS4 can take an internal medium and a surface medium extending to a specified outer radius as inputs for a single particle. Radiative transfer properties for a collection of particles can be calculated by averaging scattering and extinction cross-sections and phase matrices of particles of different materials and sizes.
        A single-scattering albedo, mean-free path length, and phase matrix are calculated for a particle describing the combined medium and are used as an input medium in SIRIS4.
        SIRIS4 is used for a multi-particle simulation by creating a large single particle with a refractive index of vacuum.
        The particles that describe a combined medium are used as diffuse scatterers inside the large vacuum particle.
        The result is effectively a multi-particle radiative transfer simulation.
        Incident rays are also launched close to the center such that they illuminate only 5\% of projected large particle surface area, so the rays encounter the spherical surface at near-normal incidence \citep{MARTIKAINEN2018, VUORI2026}.
        We use multi-particle calculations using SIRIS4 to simulate the spectral response of regolith and to calculate input parameters for RT-CB simulations.
        
        RT-CB combines radiative transfer calculations with coherent backscattering. Coherent backscattering is caused by the constructive interference of waves propagating along reciprocal paths in a multiple scattering medium.
        RT-CB calculates the light paths defined by ray tracing and constructs the electromagnetic wave traversing in the reverse direction and calculates the interference of the reciprocal light paths.
        Coherent backscattering partly induces the observed effects at small phase angles in asteroids, such as the negative polarization branch and the opposition surge, making RT-CB suitable for modeling the effects.
        We use RT-CB to simulate the polarization phase curve and geometric albedo of Phaethon.

    \subsection{Averaging over particle species}
    \label{sec:averaging}

        The scattering properties of single particles are averaged and packed to obtain a single-scattering albedo $\varpi$, a mean-free path length $\ell$, and an effective phase matrix $\bf{P}$.
        The values are used to represent an average particle in a medium consisting of the particles that were averaged and are employed in both SIRIS4 and RT-CB multi-particle simulations.
        
        The single particles for a material $i$ are defined by their radius $a_i$, as well as their scattering and extinction cross sections $\sigma^{\text{sca}}_i$ and $\sigma^{\text{ext}}_i$.
        The scattering properties of particle populations are first averaged over their size distribution, modeled using the power-law distribution $f_i(a_{ij}) \propto a_{ij}^{-\nu_i}$, where $\nu_i$ is a material specific weighting exponent defined in the modeling, with size distribution discretized into bins $j=1,\ldots,p$.
        We discretize the particle size distribution into non-uniform bins, thus, weights are of the form $w_{ij} = \int_{a_{ij}^-}^{a_{ij}^+}f_i(a_{ij})$.
        The size averaged scattering, absorption, and extinction cross sections are calculated as
        \begin{equation}
            \sigma_i = \frac{\sum_j w_{ij} \sigma_{ij}}{\sum_j w_{ij}},
        \end{equation}
        and the size-averaged scattering phase matrix $\bf{P}$ is
        \begin{equation}
            {\bf{P}}_i = \frac{\sum_j w_{ij} \sigma_{ij}^{\text{sca}} \mathbf{P}_{ij}}{\sum_j w_{ij}\sigma_{ij}^{\text{sca}}}.
        \end{equation}
        The size-averaged light scattering parameters are averaged over the material by weighing the values with the number density of the particle population $n_i = \rho_i/V_i$, where $\rho_i$ is the packing density, another multi-particle input parameter that needs to be defined, and $V_i$ is the average volume of a particle population, calculated as
        \begin{equation}
            V_i = \frac{4\pi\sum_j w_{ij} a_{ij}^3}{3\sum_j w_{ij}}.
        \end{equation}
        The final cross sections and phase matrices averaged over size and material, are;
        \begin{equation}
            \sigma = \frac{\sum_i n_i \sigma_{i}}{\sum_i n_i},
        \end{equation}
        \begin{equation}
            \textbf{P} = \frac{\sum_i n_i \sigma_{i}^{\text{sca}}\mathbf{P}_i}{\sum_i n_i \sigma_{i}^{\text{sca}}}.
        \end{equation}
        The single-scattering albedo $\varpi$ for the averaged diffuse particle in the multiple scattering simulation is now the ratio of scattering and extinction cross sections, and the mean-free path length $\ell$ is the inverse of the extinction coefficient $k^{\text{ext}}$,
        \begin{equation}
            \varpi = \frac{\sigma^{\text{sca}}}{\sigma^{\text{ext}}},
        \end{equation}
        \begin{equation}
            \ell = \frac{1}{k^{\text{ext}}}, \quad k^{\text{ext}} = \sum_i n_i\sigma_i^{\text{ext}}.
        \end{equation}

    \subsection{Input parameters}

        There are many input parameters for the regolith model, both for the single particle simulations and for the multi-particle simulations. Even though the input parameters are plentiful, they are together linked to the physical parameters of a realistic asteroid regolith, and interpretations of the asteroid surface can be made based on them.
        The main result of the modeling is not the individual input parameters of the regolith, nor should the input parameters be directly translated to Phaethon's surface.
        The input parameters are heavily dependent on each other and fundamentally on the particle refractive indices used.
        The inputs should be interpreted as trends, where the main results are the the kinds of particles that are required to model the observables.
        
        The input parameters for all of the single particles are particle shape, size, and their complex refractive index $m(\lambda) = n(\lambda) + i\kappa(\lambda)$.
        Two materials are used in the final regolith model; olivine and troilite.
        The refractive index for olivine is data of the Stubachtal olivine in the Jena database\footnote{Database of Optical Constants for Cosmic Dust, Laboratory Astrophysics Group of the AIU Jena, \url{https://www2.astro.uni-jena.de/Laboratory/OCDB/}. Olivine Stubachtal, UV-VIS-NIR, non-oriented.}, and the refractive index for troilite was derived from laboratory measurements in \citet{VUORI2026}.
        Large single particles, calculated using geometrical optics, are modeled as Gaussian-random-sphere (GRS) particles \citep{MUINONEN1996}.
        GRS is a random particle geometry model sampled from an underlying probability density function.
        A mathematical description of the GRS particles can be found in \citet{MUINONEN1996, MUINONEN2009}.
        GRS particles are defined by their relative standard deviation of radius ($\sigma_{\text{GRS}}$) and power-law index ($\nu_{\text{GRS}}$). The input values used in our modeling are $\sigma_{\text{GRS}} = 0.1$ and $\nu_{\text{GRS}} = 4$.
        Figure \ref{fig:blender_regolith} shows GRS particle shapes sampled with the input parameters used in the model.
        
        The multi-particle regolith parameters that need to be defined for each particle type are the lower and upper limits of the size distribution, the weight function of the size distribution $\nu$, and the packing density of the particles $\rho$.
        The input parameters for a medium are the single scattering albedo $\varpi$, mean-free path length $\ell$, and scattering phase matrix $\mathrm{\mathbf{P}}$.

    \subsection{Simulation workflow}

        The regolith scattering model is properly introduced in Section \ref{sec:best-fit-reg-model}, but we detail its simulation workflow here.
        We construct the regolith model from ground up by calculating the scattering properties of single particles which are averaged to generate input parameters that describe a multi-particle regolith.
        Small olivine and troilite particles are first calculated using Mie calculations and averaged over their respective size distributions to obtain effective scattering media.
        The media are used as an expressive surface texture component and as internal troilite inclusions in large olivine-hosted particles, which are simulated using SIRIS4.
        
        The second particle species is a small aggregate consisting of small troilite grains. The geometry of the aggregate is modeled with ballistic particle-cluster aggregation with 10, 20, or 30~nm troilite grains. The grain sizes are selected so that they present a single voxel in an ADDA simulation. In the final packed shape, the locations of the grains are rounded to nearest regular grid location using the grain size as the grid resolution. The equal-volume-sphere size parameters run from 0.1 to about 95 with logarithmic steps, in the number of grains this means aggregates from three up to about $89\times10^6$ voxels. For every size, three random aggregates were used, and their scattering was computed in six different orientations. Finally, the so-called incoherent scattering properties, in the form of cross-sections and phase matrices, were computed for each size over the realizations and the orientations \citep{Muinonen-2017-firstorder,Muinonen-2018-R2T2}.
        
\begin{figure*}
            \centering
            \includegraphics[width=\textwidth]{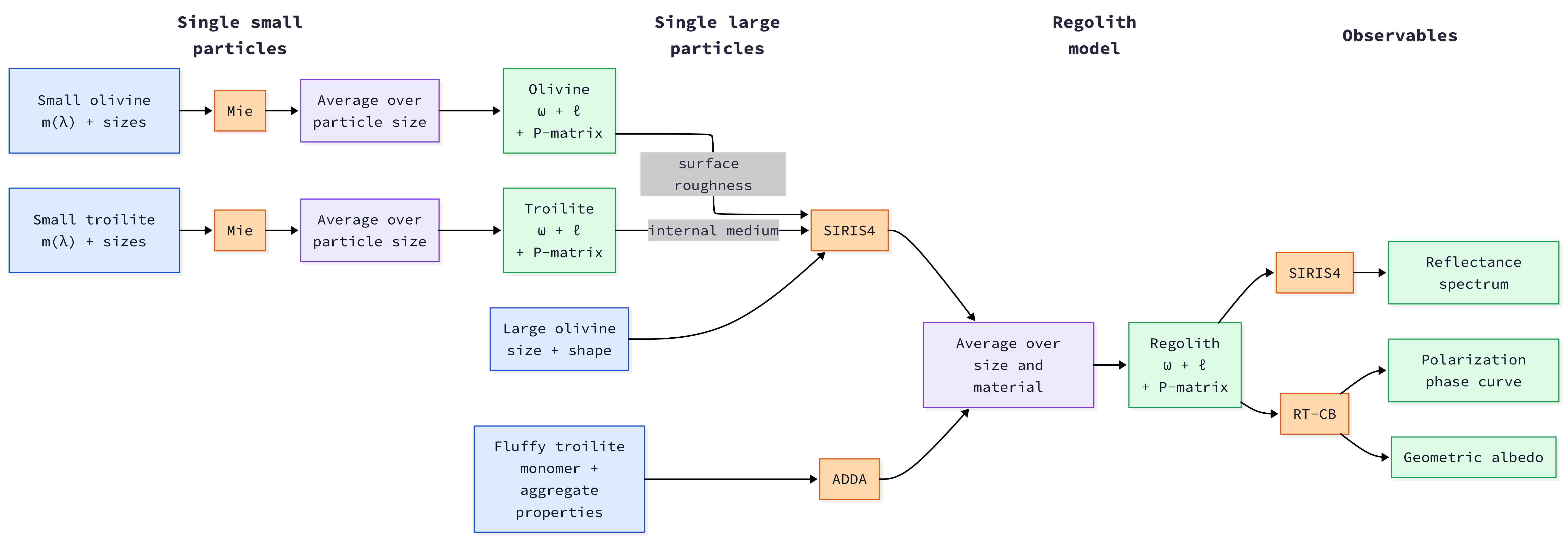}
            \caption{Flowchart of the simulation process.
            Small particles are first calculated using absMie for olivine and troilite, from which multi-particle media are created that are used as surface roughness and internal inclusions in large olivine-hosted particles. The large olivine-hosted particles are simulated using SIRIS4 and the troilite aggregate particles using ADDA. A regolith medium is created from the two large particle species. Observables are finally calculated of the regolith medium using SIRIS4 and RT-CB.}
            \label{fig:sim_flowchart}
        \end{figure*}

        The two particle species of olivine-hosted particles and troilite aggregate particles are averaged over their size distributions and further combined by averaging the particle species, to construct the effective scattering properties of the regolith.
        The final regolith medium is used to calculate the spectral response of the simulated regolith using SIRIS4, and the polarization response and geometric albedo using RT-CB.
        The simulation workflow is visualized in Figure \ref{fig:sim_flowchart}.
        
\section{Results}
\label{sec:results}

    \subsection{Best-fit regolith model}
    \label{sec:best-fit-reg-model}

\begin{figure}
            \centering
            \includegraphics[width=\linewidth]{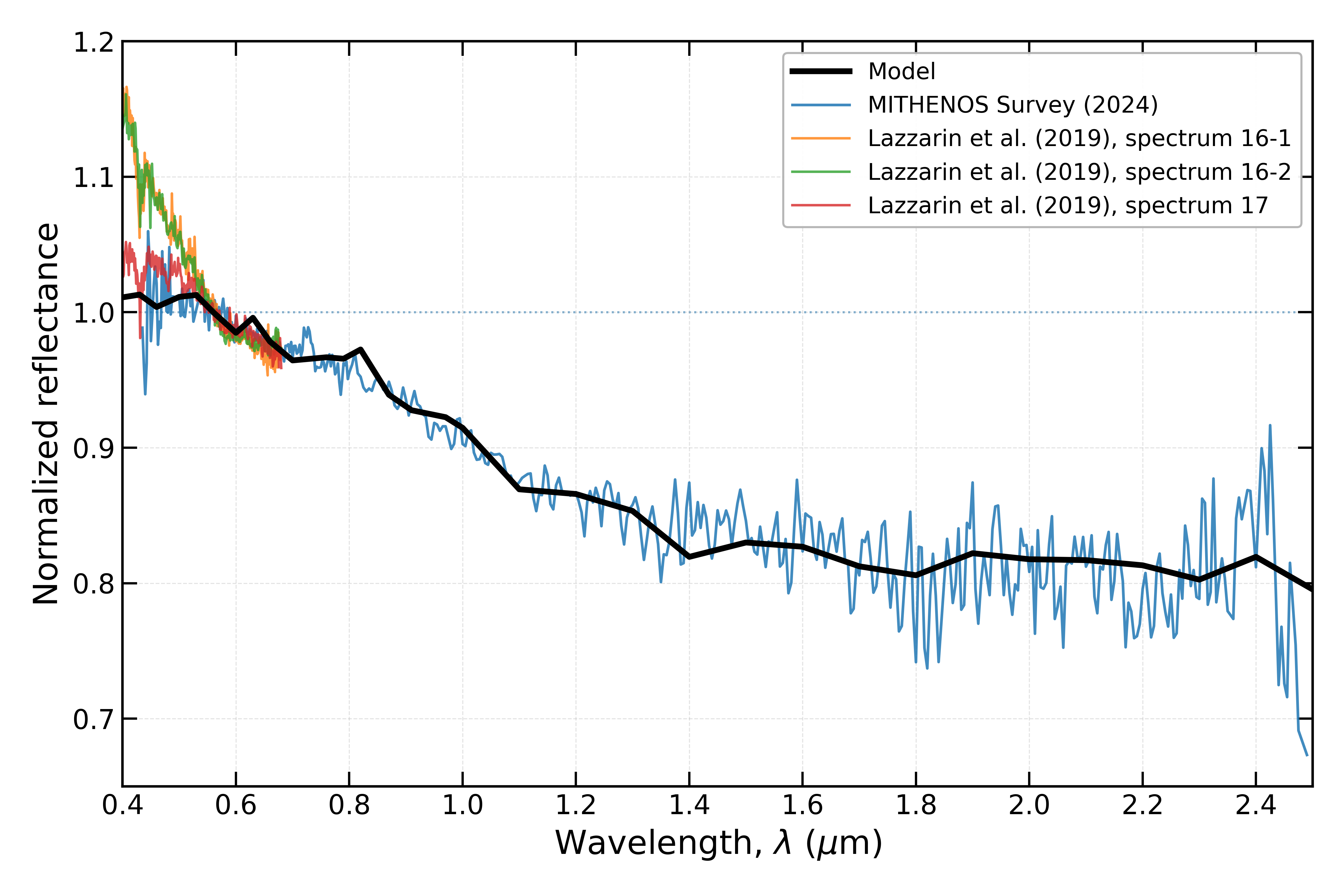}
            \caption{Predicted reflectance spectra of the best-fit model compared to observed spectra. The MITHNEOS 2024 spectrum of Phaethon was used as the comparison against which the model was constructed, as it is a recent spectral observation with a large wavelength coverage and a dense wavelength sampling. The model is accurately able to produce the spectrum in the whole Vis-NIR wavelength range.}
            \label{fig:specfit1}
        \end{figure}

\begin{figure}
            \centering
            \begin{subfigure}[H]{0.49\textwidth}
                \centering
                \includegraphics[width=\linewidth]{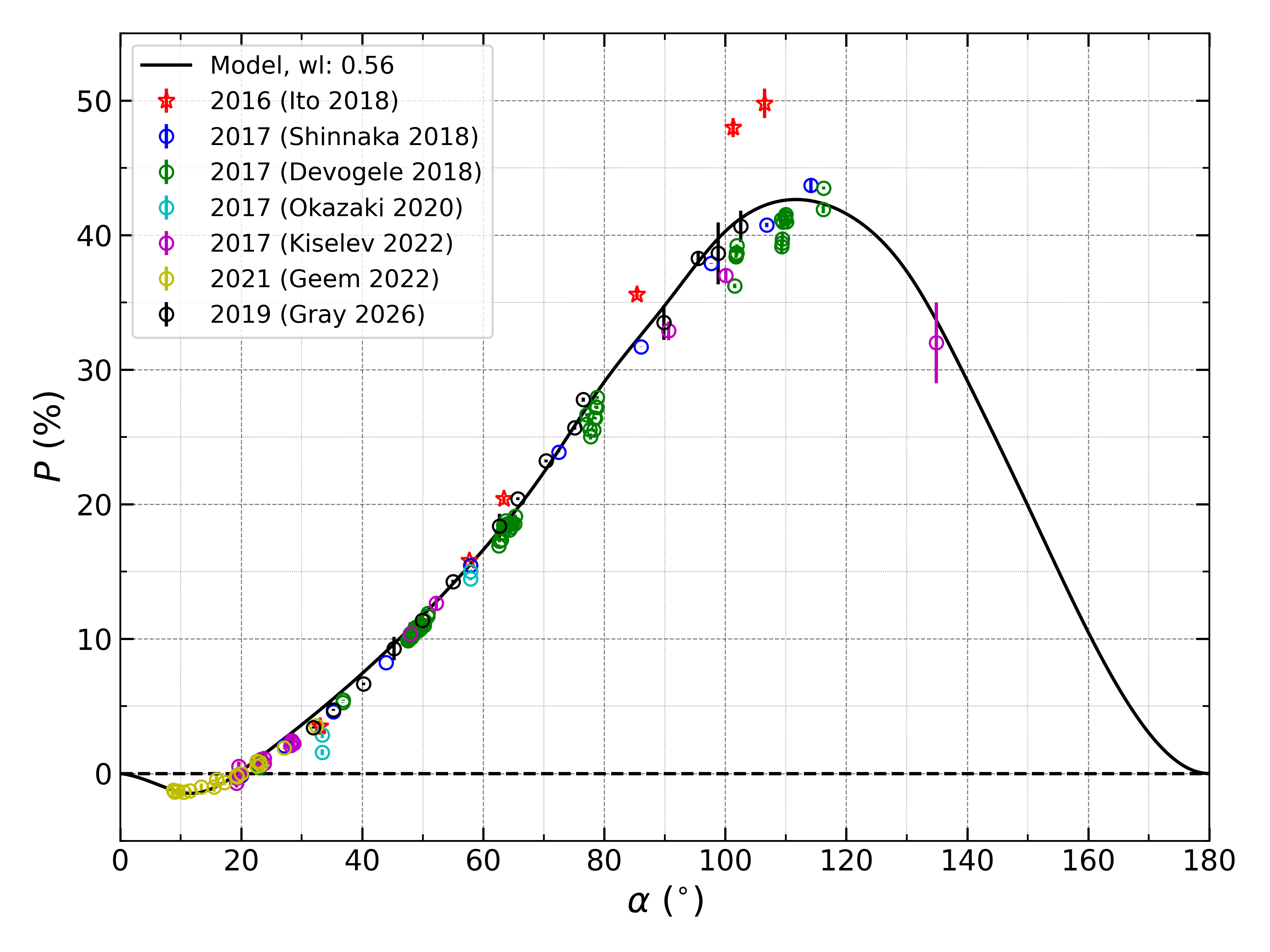}
                \caption{}
                \label{fig:polfit1_whole}
            \end{subfigure}
            \centering
            \begin{subfigure}[H]{0.49\textwidth}
                \includegraphics[width=\linewidth]{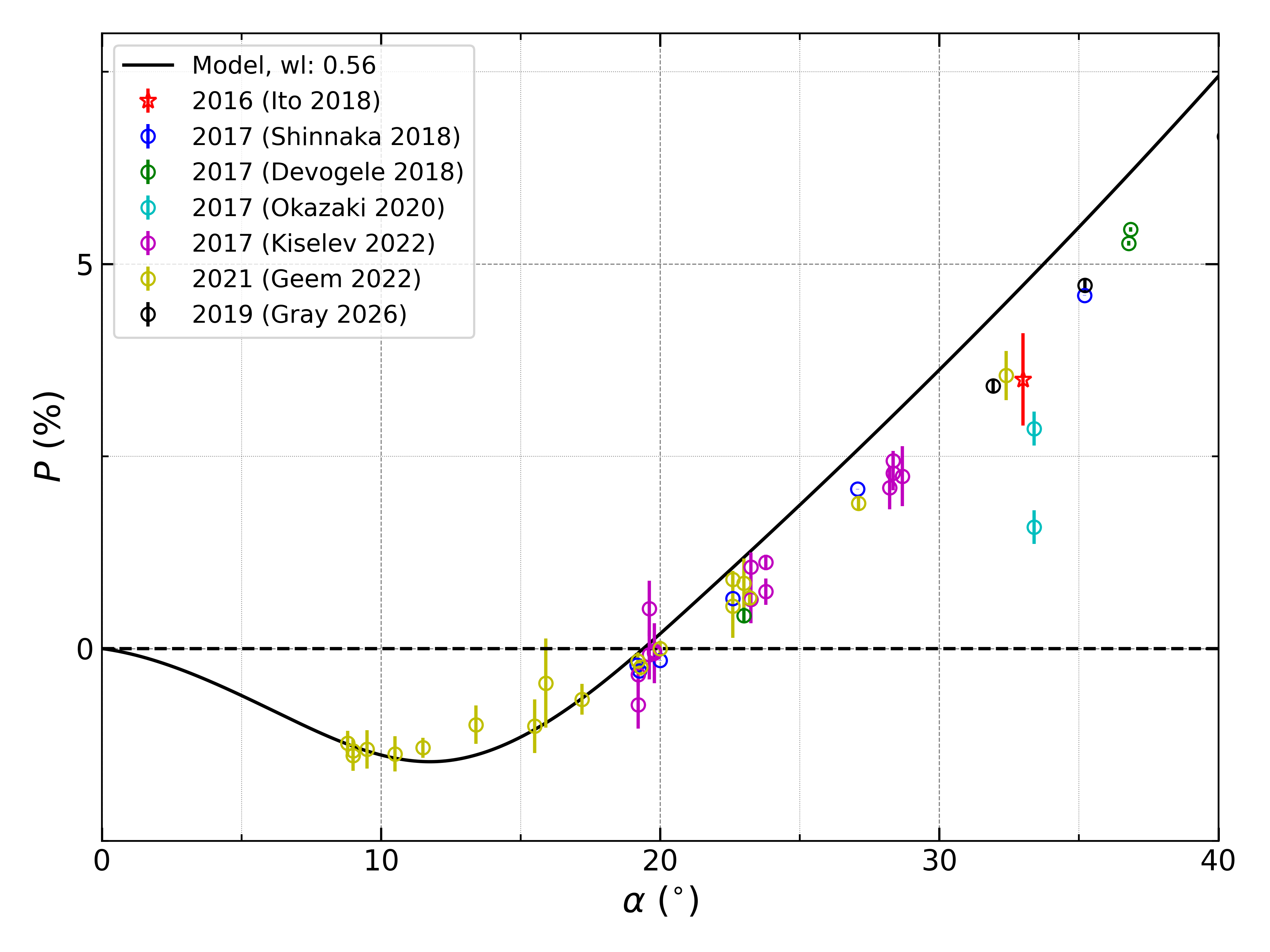}
                \caption{}
                \label{fig:polfit1_back}
            \end{subfigure}
            \caption{(a) Polarization fit of the model compared to combined polarization observations with V and R filters,
            and (b) a zoomed view of the backscattering region. The model is able to replicate well the $P_{\text{max}}$, $h$, $\alpha_0$, and $\alpha_{\text{min}}$. The angle where the maximum polarization is found, $\alpha_{\text{max}}$, is at $110^\circ$ in the model, but as the observations do not clearly show where the polarization turns downwards, the $\alpha_{\text{max}}$ for Phaethon could be found on larger angles. On the other hand, the model is consistent with the single observation at a large phase angle $\alpha \approx 135^\circ$ from \citet{KISELEV2022}.}
            \label{fig:polfit1}
        \end{figure}

        The regolith model that best describes the observations of Phaethon is constructed with two particle species, olivine-hosted particles and troilite aggregates.
        The best-fits found for the model to the spectral and polarization data of Phaethon are given in Figures \ref{fig:specfit1} and \ref{fig:polfit1}.
        The visible geometric albedo at $\lambda=0.56$~{\textmu}m result from the model is $p_{\text{V}} = 0.093$. A comparison of the model albedo to previously derived values for the albedo of Phaethon is given in Table \ref{tab:albedos}.
        A summary of the model components, rationale for their presence in the model, and the values of the multi-particle input parameters are given in Table \ref{tab:model}.
        The olivine-hosted particles are modeled as GRS olivine particles, with troilite inclusions and a mantle comprising small olivine particles mimicking the small surface texture of the particles.
        The troilite is in the form of aggregate particles comprising small tightly packed troilite monomers.
        
\begin{figure*}
            \centering
            \begin{subfigure}[H]{0.49\textwidth}
                \centering
                \includegraphics[width=\linewidth]{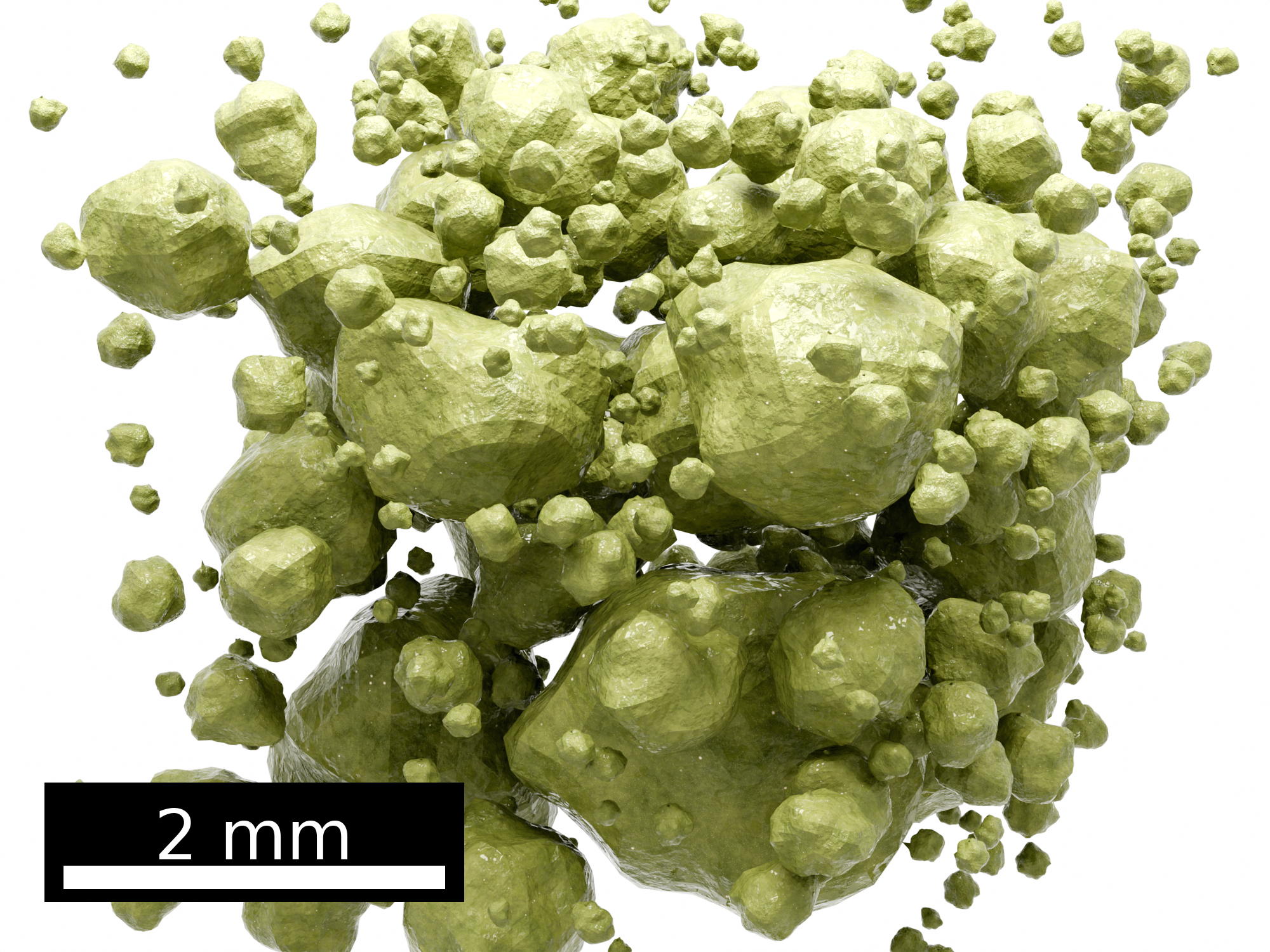}
                \caption{}
                \label{fig:blender_regolith}
            \end{subfigure}
            \begin{subfigure}[H]{0.49\textwidth}
                \centering
                \includegraphics[width=\linewidth]{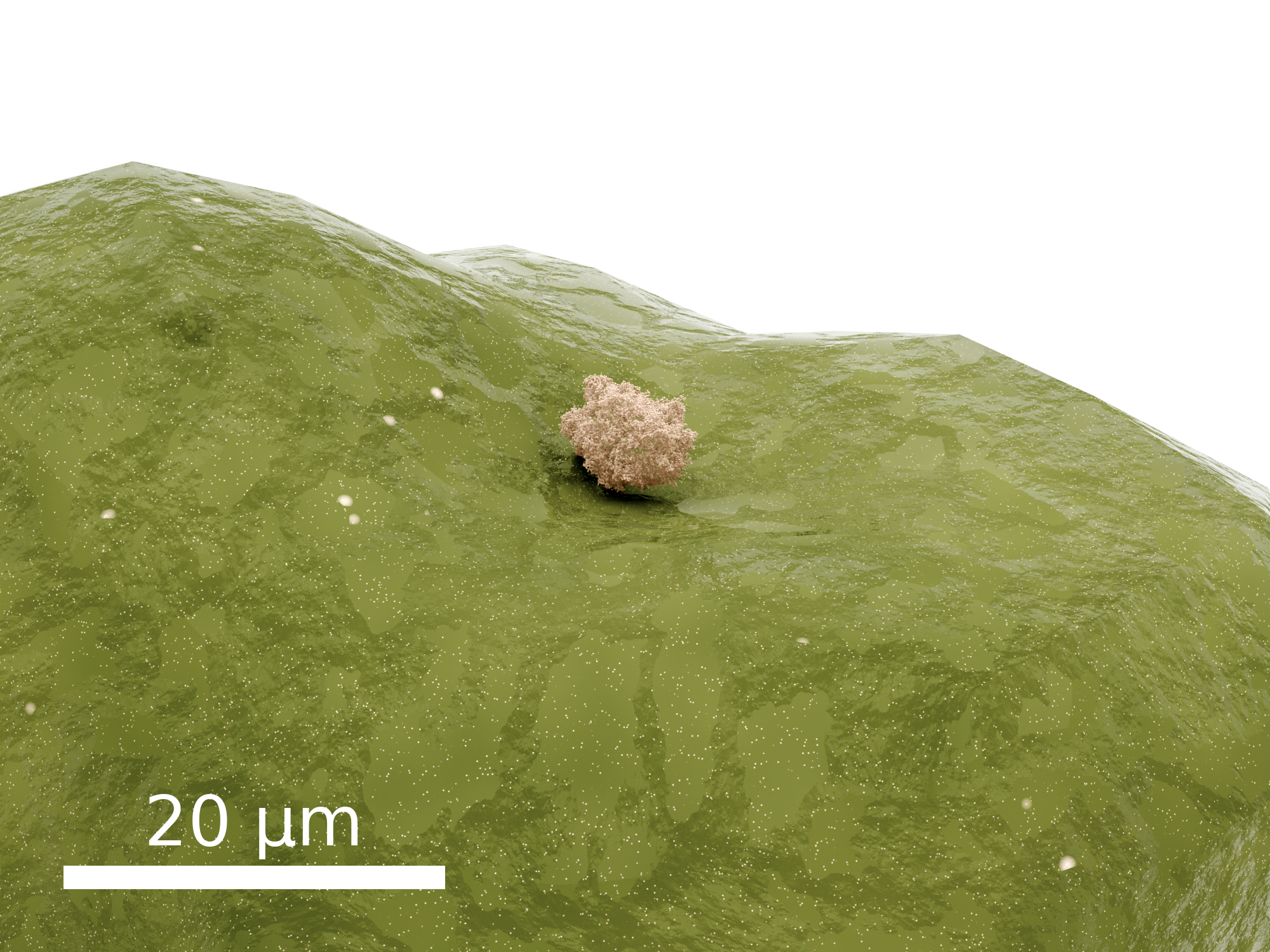}
                \caption{}
                \label{fig:blender_fluffytroilite}
            \end{subfigure}
            \caption{(a) Model of the regolith. The GRS particles with the green shade are the large olivine-hosted particles with internal inclusions and surface roughness. The particle packing density ($\rho=20$\%) and power-law size distribution parameter ($\nu=2$) represent the best-fit regolith model. The fluffy troilite aggregates are not visible due to their sizes being so small compared to olivine-hosted particles.
            (b) Closeup of a troilite aggregate particle on top of an large olivine-hosted particle. The olivine particle has very small surface texture visible, given as a height map on the surface, that is modeled as a layer of small olivine particles in the light scattering simulations, as well as brighter spots corresponding to the troilite inclusions. The figure is created in Blender software using the actual size, amount, and packing density of the particles in the model.}
            \label{fig:blender}
        \end{figure*}

        A visualization of the best-fit regolith is given in Figure \ref{fig:blender}.
        The visualization was created using Blender software, where the model regolith was constructed from 500 GRS olivine-hosted particles in the 100--2000~{\textmu}m size range, using the $\nu=2$ power law size distribution parameter and packed with a $\rho=20$\% packing density. Troilite inclusions are visualized as brighter spots on the surface, along with surface roughness represented by a displacement height map. The troilite aggregate particle was constructed using its dipole geometry representation for ADDA, with dipoles rendered as equal-sized monomers.
        Colors and illumination in the images were chosen for visual clarity, but the relative particle sizes, amounts, and densities are comparable to the input values used in our light scattering modeling.
        
        As shown by the polarization fit in Figure \ref{fig:polfit1}, the model correctly predicts an inversion angle $\alpha_0=20^\circ$, minimum degree of linear polarization $P_{\text{min}}\sim-1.5$\%, and maximum degree of linear polarization $P_{\text{max}}\sim43$\%.
        The model predicts that the phase angle corresponding to $P_{\text{max}}$ is $\alpha_{\text{max}}=110^\circ$.
        We note that $P_{\text{max}}$ is not directly sampled by the available observations, and $\alpha_{\text{max}}$ therefore remains observationally unconstrained. Thus, one could argue that the actual maximum should be located at a larger phase angle. For example, 
        earlier polarization phase curve fitting has placed the $\alpha_{\text{max}}$ around $125^\circ$ \citep{KISELEV2022, PRASAD2024}.
        Therefore, the value of $\alpha_{\text{max}}$ is the largest challenge of the model.
        In the model, the location is mainly governed indirectly by the refractive indices of olivine and troilite.
        According to our modeling, the olivine-hosted particles alone place $\alpha_{\text{max}}$ at very large phase angles, but the mixing of troilite aggregate particles, and the small surface roughness on top of the olivine-hosted particles brings $\alpha_{\text{max}}$ to lower angles. In order for the model to produce the albedo and spectral fit as well, $\alpha_{\text{max}}$ has to be confined to $110^\circ$.
        It can be argued that the olivine and troilite refractive indices used in the model are not perfect as material inputs.
        
        In light scattering modeling, we are highly restricted by the refractive indices found in literature and what we can infer ourselves.
        Even though the materials do not perfectly predict the $\alpha_{\text{max}}$, olivine and troilite still provide the best fit.
        We also tested amorphous carbon, graphite, and iron as substitutes for troilite, but out of the tested darkening agents, troilite remains the best candidate.
        It is worth noting that the two particle species model is also an idealization of the surface, as Phaethon's regolith likely comprises other materials in smaller quantities, in addition to the olivine and troilite components \citep{MACLENNAN2024, SUTTLE2024}.
        
        \begin{table}[]
        \centering
        \caption{Modeled geometric albedo at $\lambda=0.56$~{\textmu}m compared with V-band published estimates for Phaethon.}
        \begin{tabular}{lc}
        \hline
        Source & Geometric albedo, $p_{\text{V}}$ \\
        \hline
        \hline
        \textbf{This work (model result)} & $\mathbf{0.093}$ \\
        \hline
        \cite{TEDESCO2002} & $0.1066 \pm 0.011$ \\
        \cite{HANUS2016} & $0.122 \pm 0.008$ \\
        \cite{ITO2018} & $0.081$ \\
        \cite{KARETA2018} & $0.08 \pm 0.01$ \\
        \cite{SHINNAKA2018} & $0.14 \pm 0.04$ \\
        \cite{MASIERO2019} & $0.16 \pm 0.02$ \\
        \cite{KISELEV2022} & $0.060 \pm 0.005$ \\
        \cite{GEEM2022} & $0.11 \pm 0.02$ \\
        \cite{GRAY2026} & $0.094$--$0.12$ \\
        \hline
        \end{tabular}
        \label{tab:albedos}
        \end{table}
        
        \begin{table*}
            \centering
            \caption{Summary of the best-fit regolith-model components and the
            observational constraints motivating their inclusion. Here, $S$, $P$,
            and $p_{\text{V}}$ denote the spectrum, polarization phase curve, and geometric
            albedo.
            The model has two main particle species that are mixed together to produce the regolith; olivine-hosted particles (denoted with \textbf{O}), and troilite aggregate particles (\textbf{T}).
            \label{tab:model}}
        
            \small
            \renewcommand{\arraystretch}{1.25}
        
            \begin{tabular}{@{}llll@{}}
                \tableline
        
                \parbox[t]{0.04\textwidth}{\textbf{ID}}
                &
                \parbox[t]{0.27\textwidth}{\raggedright
                \textbf{Model component}}
                &
                \parbox[t]{0.16\textwidth}{\raggedright
                \textbf{Main constraints}}
                &
                \parbox[t]{0.45\textwidth}{\raggedright
                \textbf{Rationale for inclusion in the model}}
                \\
        
                \tableline
        
                \parbox[t]{0.04\textwidth}{\textbf{O1}}
                &
                \parbox[t]{0.27\textwidth}{\raggedright
                \textbf{Large olivine particles} --- Gaussian-random-sphere particles
                with sizes of $100$--$2000$~{\textmu}m.}
                &
                \parbox[t]{0.10\textwidth}{\raggedright
                $p_{\text{V}}$, $P$}
                &
                \parbox[t]{0.45\textwidth}{\raggedright
                Provides the large-scale geometric-optics behavior of the regolith.
                Smaller particles would produce a red spectrum,
                other bright silicates behave optically similarly to olivine.}
                \\
        
                \parbox[t]{0.04\textwidth}{\textbf{O2}}
                &
                \parbox[t]{0.27\textwidth}{\raggedright
                \textbf{Troilite inclusions} --- embedded within the \textbf{O1} olivine particles, with sizes of $0.01$--$1$~{\textmu}m.}
                &
                \parbox[t]{0.10\textwidth}{\raggedright
                $p_{\text{V}}$, $P_{\max}$}
                &
                \parbox[t]{0.45\textwidth}{\raggedright
                Makes the large olivine particles opaque, lowers the
                albedo, and increases $P_{\max}$. Inclusion-free
                olivine cannot reproduce both constraints simultaneously.}
                \\
        
                \parbox[t]{0.04\textwidth}{\textbf{O3}}
                &
                \parbox[t]{0.27\textwidth}{\raggedright
                \textbf{Micron-scale olivine particles} --- as surface texture on the \textbf{O1} olivine particles, with sizes of $0.01$--$1$~{\textmu}m.}
                &
                \parbox[t]{0.10\textwidth}{\raggedright
                 $P_{\min}$, $p_{\text{V}}$}
                &
                \parbox[t]{0.45\textwidth}{\raggedright
                Produces the negative-polarization branch and slightly increases
                the albedo. Smooth olivine particles do not reproduce the observed
                backscattering behaviour.}
                \\
        
                \parbox[t]{0.04\textwidth}{\textbf{T1}}
                &
                \parbox[t]{0.27\textwidth}{\raggedright
                \textbf{Fluffy troilite aggregates} --- with sizes of
                $0.04$--$6$~{\textmu}m.}
                &
                \parbox[t]{0.10\textwidth}{\raggedright
                $S$, $P_{\max}$, $p_{\text{V}}$}
                &
                \parbox[t]{0.45\textwidth}{\raggedright
                Fine absorbing particles produce the blue spectral slope and slightly reduce
                $P_{\max}$. Spherical Mie particles produce too low albedo. Larger troilite particles have little
                effect on the modeled observables. The tested alternative darkening
                agents do not reproduce the spectrum and polarization
                simultaneously.}
                \\
        
                \tableline
            \end{tabular}
        
            \vspace{0.5em}
        
            \parbox{0.90\textwidth}{\footnotesize
            The multi-particle input parameters; packing densities ($\rho$) and size-distribution exponents ($\nu$) are:
            $\rho_{\mathrm{O1}}=0.2$, $\nu_{\mathrm{O1}}=2$,
            $\rho_{\mathrm{O2}}=0.02$, $\nu_{\mathrm{O2}}=4$,
            $\rho_{\mathrm{O3}}=0.04$, $\nu_{\mathrm{O3}}=4$,
            $\rho_{\mathrm{T1}}=0.004$, and $\nu_{\mathrm{T1}}=5.3$.}
        \end{table*}
        
        We constructed the regolith model by adding the necessary components to model all spectral and polarization effects from observations.
        It needs to be stated that, while the model is as simple as possible, there are a large number of input parameters.
        Combined with the computational cost of the simulations, which can be in the order of CPU-years, this makes using rigorous methods to find the best fit input parameters and their uncertainties, such as Markov chain Monte Carlo, computationally impractical.
        
        In \citet{PENTIKAINEN26} it is shown that to produce the high $P_{\max}$ typical for C- and B-type asteroids, large absorbing particles producing mainly external reflections are needed.
        We make an assumption in the modeling that Phaethon's regolith is mainly composed of olivine. The olivine used here ($\kappa\sim10^{-6}$) is by itself too bright to produce the required $P_{\max}$, so dark troilite inclusions are added to the olivine matrix to increase the absorption in the material.
        The large olivine-hosted particles alone are able to produce a low-albedo surface with a relatively flat spectrum that is red sloped, i.e., albedo increases with wavelength.
        To produce the blue spectrum, small troilite aggregate particles are mixed in with the larger olivine-hosted particles.
        The amount of troilite needed to blue the spectra also lowers the albedo considerably.
        To add enough troilite for the blue spectra while preserving the albedo and high polarization, the troilite is modeled as fluffy aggregate particles.
        In such a dark regolith, the negative polarization branch is only governed by the smallest bright particles.
        Small olivine is thus added to the surface of the olivine-hosted particles, which simulates surface roughness.
        The small surface roughness olivine particles are able to reproduce $P_{\min}$, but by themselves do not affect the rest of the polarization curve, or the spectrum.

    \subsection{Fluffy aggregate particles}

        \citet{KURODA2018} examine the polarization response of (152679) 1998 KU$_2$, which is an extremely dark asteroid with a geometric albedo of 0.018--0.030 and a high $P_{\max} \sim50$\%.
        The $P_{\max}$ is significantly higher than other dark solar system objects usually display.
        To explain the high $P_{\max}$, they conclude that 1998 KU$_2$ surface should be covered in micron-sized fluffy aggregates composed of carbon nanoparticles. The dark fluffy aggregates contribute to sharp increase in $P_{\max}$ caused by multiple scattering of nanoparticles which also keeps the albedo low \citep{VOLTEN2007, KOLOKOLOVA2010, TAZAKI2022}. \citet{KURODA2018} compare polarization response of 1998 KU$_2$ to Phaethon, which displays similar polarization and albedo effects. We think that their interpretation is correct, and that Phaethon's polarization is governed by a combination of external reflections from large opaque particles and multiple scattering from nanoparticles.
        
        The blue spectrum and $p_V\sim0.1$ can be explained by using spherical Mie particles of troilite in the model. However, using Mie particles, the $P_{\max}$ is only 35\%. Similarly, the blue spectral response and $P_{\max} \sim 45\%$ could be explained using Mie particles, but the resulting $p_V=0.06$.
        Changing the troilite Mie particles to fluffy aggregates, as proposed in \cite{KURODA2018}, all three---blue spectrum, $P_{\max} \sim 45\%$, and $p_V=0.1$---can be achieved.

    \subsection{Blue spectrum}

        The blue spectral slope is produced by mixing large (100--2000~{\textmu}m) and small particles (0.04--6~{\textmu}m).
        Specifically, there needs to be a deficiency of particles that have intermediate sizes.
        These intermediate particles do not necessarily need to be completely absent, but their contribution to scattering processes needs to be lower than that of the very largest and smallest particles.
        The blueing effect of mixing a small darkening agent and large brighter material has been noted earlier \citep{CLOUTIS1990, GALIANO2020, SULTANA2021, LOEFFLER2022}.

        \citet{LISSE2022} propose that the blue spectrum of Phaethon could be caused by a loss of iron which usually reddens a spectrum.
        However, in \citet{LOEFFLER2022}, it is found that the blueing effect caused by mixing is seen with different darkening agents, even with some that are by themselves very red.
        Based on modeling presented here, we suspect that blueing is caused by the smallest dark particles.
        With only large and small particles present, the effect that the small dark particles have on light scattering is considerably higher than the largest particles, and the large particles do not contribute considerably to spectral shape.
        The cause for the blueing has been proposed to be Rayleigh scattering \citep{CLARK2008, BROWN2014, SCHRODER2021}, although \citet{LOEFFLER2022} suspect that the Rayleigh scattering is not the cause, as their samples are not in the Rayleigh regime.
        Further study on the regolith particle size dependency that causes the blue spectral slope is out of the scope of this article, but it is something that should be studied further.

    \subsection{Explanation of surface heterogeneity}

        Our regolith model is able to reproduce the observations from Phaethon, and can therefore be used to study the possible surface heterogeneity.
        The more extreme cases of different observations, such as spectral data from \cite{LAZZARIN2019} and polarization data from \cite{ITO2018}, will now be addressed with the model.
        We attempt fitting the different spectra and polarization observations by tuning all the parameters in the model and investigating how they each affect the results.
        We assumed that a single parameter can be found in the model explaining the differences seen both in spectra and polarization.
        However, uniquely coupling spectral and polarization differences to the same source for the heterogeneity is not possible, as model parameters can be found that reproduce the heterogeneous spectral data, while keeping the polarization unchanged. The same is true for the more extreme case of polarization.
        Therefore, we detail the model parameters that change spectra and polarization separately.
        The implications of the results are analyzed in detail in Section \ref{sec:heterogenity}.

        \subsubsection{Spectrum}

\begin{figure}
                \centering
                \includegraphics[width=\linewidth]{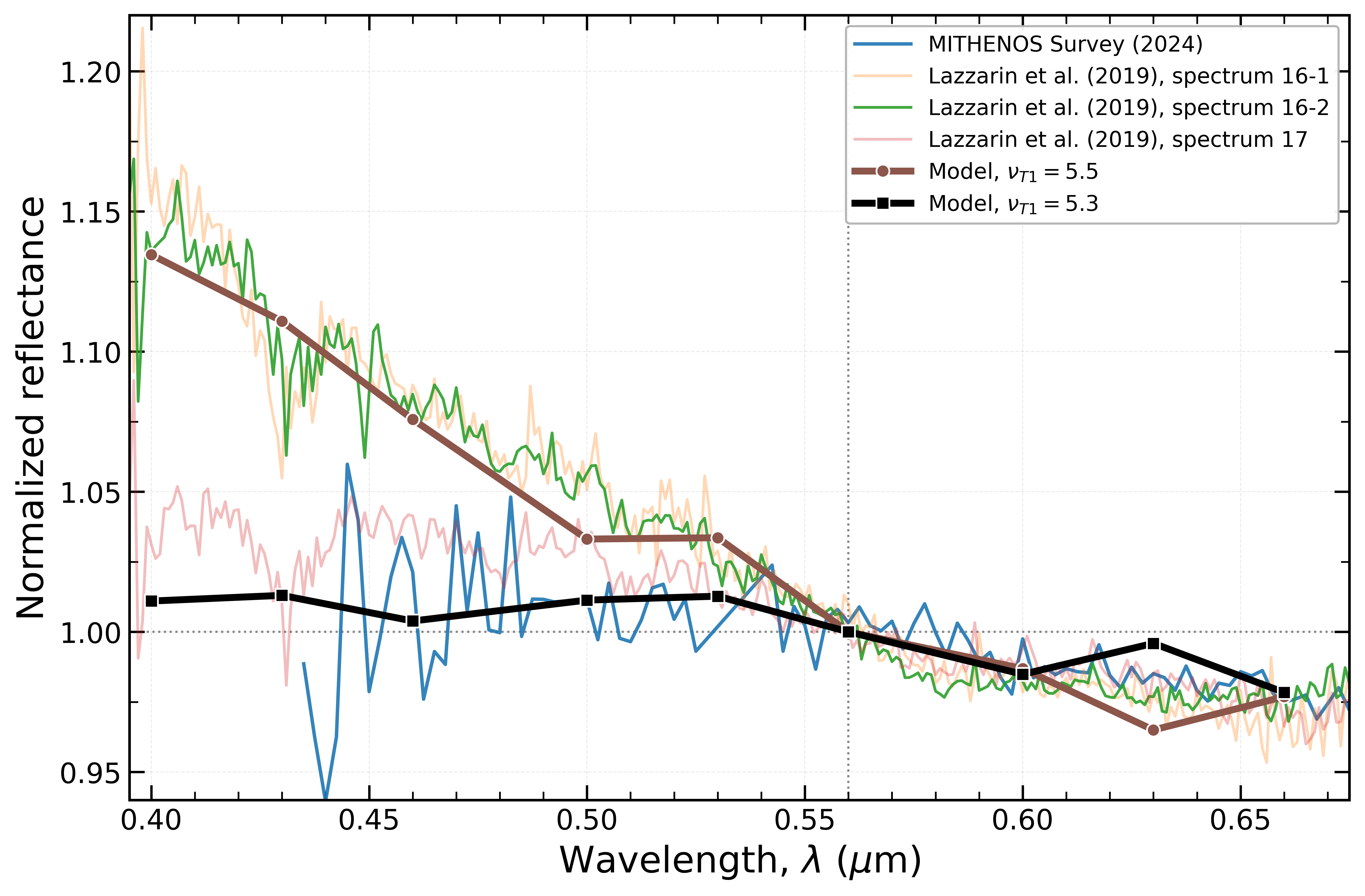}
                \caption{Comparison of Vis-spectral results from the original model with troilite aggregate size distribution exponent of $\nu_{\text{T1}}=5.3$ and $\nu_{\text{T1}}=5.5$. The smaller exponent value of the weight function accurately models the bluer spectra (Spectrum 16-1 and 16-2) from \citet{LAZZARIN2019}, and physically translates to a larger abundance of smaller troilite particles between the regoliths. All other spectral slopes that can be found between the two simulation results could also be modeled by changing the exponent value.}
                \label{fig:specfit2}
            \end{figure}

            According to our modeling, the differences seen in the spectral data in UV and visible wavelengths can be explained by changing the size distribution of the troilite aggregate particles in the model from $\nu_{\text{T1}}=5.3$ to $\nu_{\text{T1}}=5.5$.
            The change in size distribution from $\nu_{\text{T1}}=5.3$ to $\nu_{\text{T1}}=5.5$ means that the contribution to light scattering from the troilite aggregate particles is coming more from the smallest aggregates.
            The difference in the model parameters between the extreme cases of spectra, thus,  are not large.
            The models that fit the two extreme spectral cases both give the same polarization phase curve and albedo.
            Spectral fits of the two models with different exponent values compared to spectral observations in the visible wavelength range are given in Figure \ref{fig:specfit2}.
            
        \subsubsection{Polarization}

\begin{figure}
                \centering
                \includegraphics[width=\linewidth]{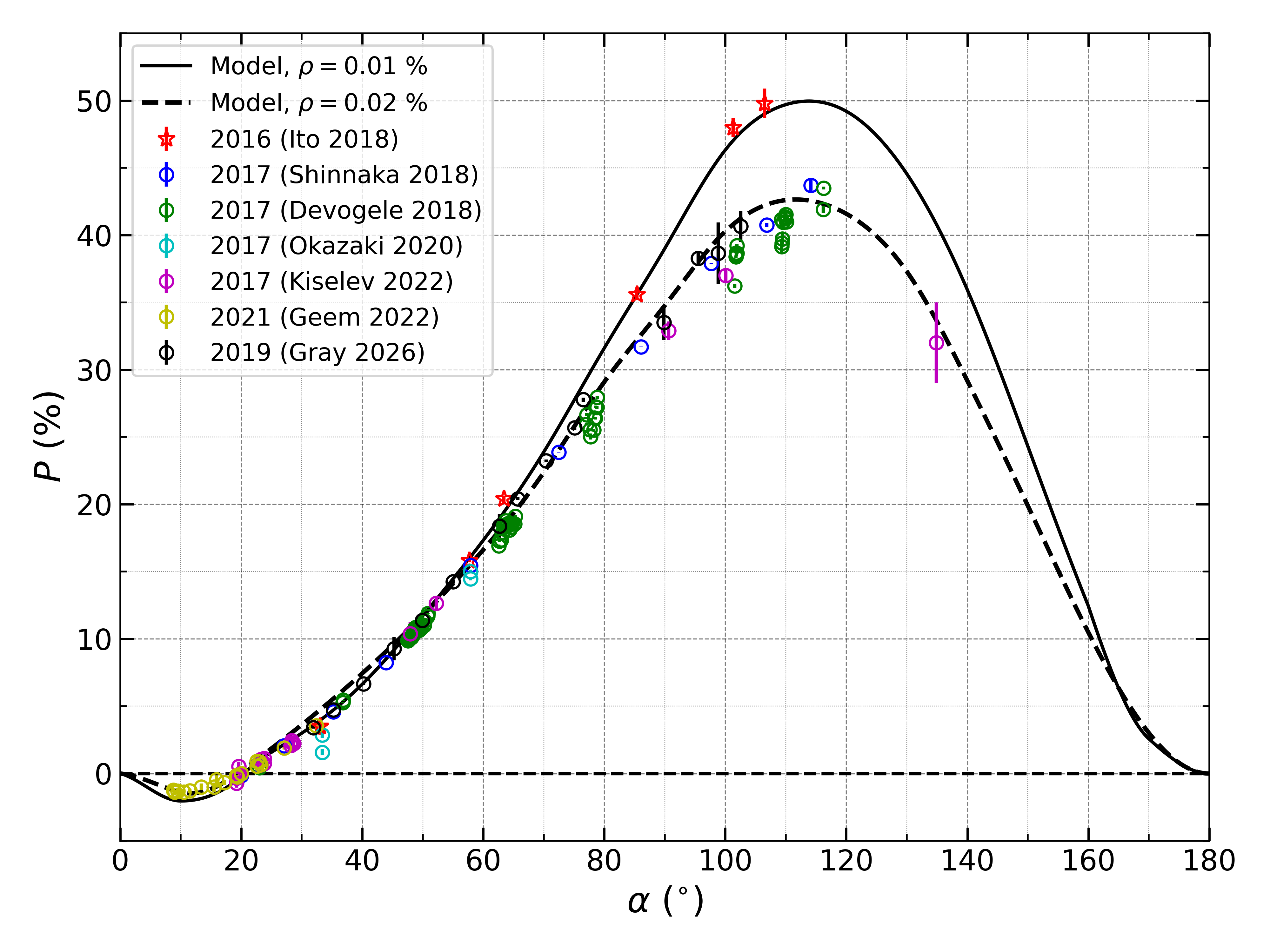}
                \caption{Polarization result for a model where the packing density of troilite aggregate particles has been halved from 0.02\% of the original model to 0.01\%. By halving the amount of troilite aggregates, a good fit to the higher polarization observations of \citet{ITO2018} is achieved.}
                \label{fig:polfit2}
            \end{figure}

            The extreme polarization found in \cite{ITO2018} can be modeled by halving the amount of troilite aggregate particles in the regolith model from 0.02\% to 0.01\% ($\rho_{\text{T1}} = 0.004$ $\rightarrow$ $\rho_{\text{T1}} = 0.002$).
            The smaller amount of troilite aggregates allows the large olivine-hosted particles to contribute more to the scattering.
            The olivine-hosted particles are very dark, and the larger contribution from them raises $P_{\max}$ and moves $\alpha_{\max}$ to larger phase angles.
            The minimum polarization branch also deepens, as the surface roughness on top of the olivine-hosted particles contributes more to the polarization response.
            The geometric albedo further decreases from 0.09 to 0.07.
            A comparison of the two polarization results of the original model and the model with half the amount of troilite is given in Figure \ref{fig:polfit2}.
            
\section{Discussion}

    \subsection{Cause of heterogeneity}
    \label{sec:heterogenity}

        The regolith modeling shows that the observed spectral and polarimetric differences do not require a common physical mechanism.
        The spectral differences can be modeled by changing the size distribution of the smallest troilite aggregates, whereas the high polarization measured by \citet{ITO2018} requires a change in their relative contribution.
        The normalization of low-albedo spectra might visually give the impression that the changes in the spectra are dramatic, but based on our modeling, the differences in the observed spectra are caused by small differences in the sizes of the smallest regolith particles ($\nu_{\text{T1}}=5.3 \rightarrow \nu_{\text{T1}}=5.5$).
        Such small variability is probably present all over the asteroid surface, and there is no reason to assume that the smallest particles in the regolith would be completely homogeneous when the asteroid is exposed to heating and space weathering.
        For the above reasons, we believe that the UV-Vis spectral range is not a good indicator of surface heterogeneity over the Phaethon's surface.
        In contrast, the polarization phase curve provides a stronger constraint on the physical differences in the regolith.
        
        \citep{GRAY2026} investigated how the possible surface heterogeneity can be inferred from observations.
        They found that the higher polarization maxima observed in 2016 by \citet{ITO2018} and in their new 2019 observations could be explained by the observations sampling the northern hemisphere.
        Based on flux distribution calculations, the 2016 observations only sample the northern hemisphere, whereas 2019 also has some contribution from the southern hemisphere, but still being mostly weighted towards the North.
        The incidence and emission angles of the observations also have different trends for the 2016 observations.
        The 2017 and 2019 disk-integrated observations could have more pronounced shadowing and masking effects than the 2016 observations.
        Shadowing occurs when particles block incident light from reaching other particles, while masking occurs when particles obstruct the scattered light from reaching the observer.
        Such effects are expected to lower the polarization at high phase angles, as certain externally reflected light paths become impossible.
        The authors suggested both hemispherical heterogeneity and different observation geometries as potential explanations for the intermediate 2019 polarization response, with northern hemisphere raising $P_{\text{max}}$ and observation geometry lowering $P_{\text{max}}$, but were unable to definitely infer the cause, as the light scattering response of Phaethon is dependent on the properties of the surface regolith.
        
        Light scattering modeling in this article suggests that the 2016 higher $P_{\text{max}}$ observations can be explained by halving the amount of the troilite aggregate particles, which are the small particle component in the model.
        The small troilite aggregates lower the $P_{\text{max}}$ of the regolith, compared to a regolith only made of the olivine-hosted particles.
        The underlying mechanism in halving the troilite amount is, that the olivine-hosted particles start contributing more to the polarization, raising $P_{\text{max}}$, shifting it to slightly higher $\alpha_{\text{max}}$, and lowering albedo from 9\% to 7\%.
        If the modeled reduction in the small-particle population represents a true regolith difference on Phaethon, the regolith with a high polarization response would also be expected to have a lower reflectance.

        \citet{MACLENNAN2022} found heterogeneity using thermal inertia modeling between northern and southern latitudes of Phaethon with the South having smaller ($<300$~{\textmu}m) particles and the North larger ($>5$~mm) particles.
        Modeling methods in thermal and Vis-NIR are differently sensitive to the regolith properties.
        For example, thermal modeling is expected to be more sensitive to larger single particles than our light scattering modeling.
        Therefore, in our analysis, we do not interpret the particle size differences inferred in \citet{MACLENNAN2022} precisely but as representative of size distribution changes in the regolith particles.
        Nevertheless, the regolith properties inferred using different methods are complementary, and the particle size differences from thermal modeling could provide a straightforward explanation to the polarization response between the hemispheres, as larger particles are expected to cause a higher $P_{\text{max}}$ \citep{FRATTIN2019, MUNOZ2021}.
        However, moderate changes in the size distribution of the olivine-hosted particles in the model do not affect \(P_{\max}\), because the particles are in the geometric-optics regime and are optically opaque at the modeled wavelengths. Their single-particle polarization response is effectively independent of size. Even when the largest particles are replaced with smaller particles, \(P_{\max}\) increases rather than decreases, because the remaining smaller particles of the olivine-hosted population contribute more strongly to the scattering response of the regolith.
        Our modeling does not support the notion that size differences in $> 100$~{\textmu}m particles, such as those found in \citet{MACLENNAN2022} are directly responsible for the higher $P_{\text{max}}$ in the 2016 observations.
        
        Even though the grain size cannot explain the polarization difference directly, it could still be responsible indirectly.
        If the northern and southern latitudes have different regolith particle sizes, they would create different local topography through the regolith structure.
        Depending on how the local topography is constructed, some of the particles might also be obscuring others.
        For example, the smaller troilite aggregate particles could be more hidden on the northern hemisphere, as the olivine-hosted particles size are increased.
        In the statistically mixed regolith model, a physical reduction in the abundance of small troilite aggregates is indistinguishable from any process that reduces their effective contribution to the observed scattered light.
        Using our modeling, we cannot infer if the local topographies can produce the different polarization curves, but we are able to give a physical mechanism that can be responsible for such differences.
        Halving the amount of the troilite aggregate particles in the model, lowers their effective contribution to light scattering compared to the larger olivine-hosted particles, and a higher effective contribution of the olivine-hosted particles raises $P_{\text{max}}$.
        A regolith with a smaller amount of troilite aggregates is optically equivalent to one where their effective contribution is smaller.
        
        The local topography change is also consistent with the observation-geometry effect hypothesized in \cite{GRAY2026}.
        Even though the smaller dark particles would be more hidden on the northern hemisphere, with large incidence, emission, and phase angles, the contribution of multiple scattering is more pronounced and the polarizing effect of the northern hemisphere is lowered.
        Such multiple scattering could also start probing the otherwise hidden troilite aggregates, which further lowers $P_{\max}$.
        
        Using the modeling in this article, we cannot infer how certain particles would shadow and mask each other, and what are the possible first-order and multiple scattering contributions from the different regolith species in a given observation geometry.
        The regolith model provides a physical mechanism capable of reproducing the higher 2016 polarization, and is consistent with both intrinsic regolith heterogeneity and an observation-geometry-dependent change in the effective contribution of the particle populations hypothesized in \cite{GRAY2026}.
        To physically infer the magnitudes of the two effects, the regolith model should be extended to a topography model with an appropriate shadowing function to study how the local topography can change polarization. Such modeling remains outside the scope of this work.

    \subsection{Phaethon's meteorite analogue}

        A clear distinction between CY- and CM-meteorites is the iron sulphide abundance. CMs contain $\sim2$\% sulphides where CYs have $\sim20$\% \citep{KING2019, KIMURA2020}.
        Our model uses only 0.2~vol\% troilite aggregate particles mixed in with olivine-hosted particles, to achieve the dark blue spectrum, favoring CMs as the correct analogue.
        Clear distinction must be made, that inside the olivine-hosted particles, the amount of troilite inclusions is 2~vol\%, which is also the lower limit. Raising the troilite amount inside the large olivine-hosted particles does not change their scattering properties once the olivine-hosted particles become opaque.
        It is a fair assumption that the amount of troilite inside the olivine should not be close to 20~vol\%. For the modeling to work, surface reflections of large particles need to be dominated by olivine. The amount of troilite inside the olivine should be expected to be the same amount that is found on the surface of the olivine-hosted particles. With 20~vol\%, the surface itself starts acting closer to a troilite particle than an olivine particle. If there is 20~vol\% troilite inside the olivine-hosted particles, it should be found only inside the particle and not close the surface.
        Next, we will strictly be talking about the amount of troilite aggregates mixed with the olivine-hosted particles, not the inclusions.
        
        We construct the regolith model by fitting spectral, polarization, and albedo data at the same time. Multiple observables provide heavy constraints to which the model needs to adhere.
        Troilite is an efficient darkening agent, and even a small amount is enough to blue the spectrum.
        If the amount of troilite is raised from 0.2~vol\%, the spectral shape starts to flatten from blue. To keep the blue spectral shape, the size distribution parameter $\nu$ of troilite aggregate particles needs to be raised.
        In other words, to produce the spectral shape with a larger quantity of troilite, the troilite needs to be smaller.
        So even though the spectral shape can provide lower limits for the amount of the darkening agent, it gives no clear upper limit by itself.
        
        The polarization curve also depends on the size of the particles.
        When smaller troilite aggregate particles are used, the polarization fit suffers as a consequence.
        The $P_{\text{min}}$, $\alpha_{\text{min}}$, $\alpha_0$, $P_{\text{max}}$, and $\alpha_{\text{max}}$ can be modeled correctly even with smaller troilite aggregates.
        The problems are visible at phase angles between $\alpha_0$ and $\alpha_{\text{max}}$, where the slope of the polarization curve becomes too steep, no longer aligning with the observations.
        The shape of the polarization phase curve works as a limit in the model on the size of troilite aggregates, and combined with the blue spectral slope, as a limit for the amount.
        
        With a larger amount of troilite aggregate particles the albedo also decreases, although the effect is not straightforward.
        To understand why a small shift in volume fraction can cause differences in light scattering, the mineral abundances should be translated to extinction cross sections. The very dark troilite contributes to 96\% of the regolith optical cross section although the amount of small troilite aggregates in the model is just 0.2~vol\%. With 20~vol\% troilite aggregate particles, the amount of contribution to light scattering through the extinction cross section is practically 100\%, and the albedo decreases.
        On the other hand, adding smaller particles when increasing the troilite aggregate particle amount, there is more contribution coming from multiple scattering of smaller particles which generally raises albedo.
        So the overall decrease of albedo is not linear, and not a good indicator on the amount of troilite.
        
        For the above reasons, the requirement for a small abundance of troilite favors CM as the more appropriate meteorite analogue for Phaethon over CY.
        To make a strong case for CM, the assumption would need to be made that the optically effective surface components we are modeling are representative of the true composition of Phaethon.
        
        The observed scattered light in the Vis-NIR range samples at most the first few millimeters of Phaethon's surface. There is, in general, no reason to suspect that an asteroid's surface layer, which is ravaged by space weathering, would fully reflect its interior composition \citep[e.g.][]{MATSUOKA2023, NOGUCHI2023}. This is especially true for Phaethon, which has a perihelion distance $q=0.14\,$au \citep{TAKIR2020, SUTTLE2024}. 
        The troilite could very well be found as large inclusions inside the asteroid, as is seen in CY meteorite samples \citep{SUTTLE2023}. The larger inclusions, even if present, do not affect our modeling results in any way, as the olivine-hosted particles are already opaque in the studied wavelengths.

    \subsection{Heated CM as an analogue for Phaethon's regolith}

        More support for the CM analogue can be found by comparing our regolith model to heated fine-grained CM simulants \citep{GEEM2026}.
        Heating the CM simulant removes large-scale roughness, but preserves the smaller scale surface roughness. The surface roughness on the olivine-hosted particles, required by our model, is thus retained even in intensely heated regolith.
        An even more interesting finding is what happens to the small structure in the heated sample:
        \citet{GEEM2026} note that due to heating, small particles smoothen and consolidate together to produce particles that structurally resemble the troilite aggregate particles in our model.
        Even though the particles used in our model do not compositionally match the smallest particles in the heated simulants, the heating experiments do show evidence for an effect that could create the fluffy particle structures on Phaethon.

    \subsection{Implications for the DESTINY$^+$ mission}
        DESTINY$^+$ will image Phaethon with multiband photometry in the visible wavelength range using MCAP (Multiband CAmera for Phaethon).
        As our modeling has showed, changes in spectra in the MCAP wavelength range (0.39--0.85~{\textmu}m) can be caused by small size differences of the regolith particles.
        We expect that color differences in the MCAP observations will be more representative of the particle sizes than the regolith composition.
        
        DESTINY$^+$ is also imaging Phaethon's surface in a spatially resolved manner and could be able to study the surface heterogeneity.
        In our model, the polarization differences are caused by relative abundance of the small troilite aggregate particles.
        The model also predicts that there should be an albedo difference between the two polarization signals.
        The photometric observations by DESTINY$^+$ can then be used to identify regions with systematically lower reflectance, which could correlate with the higher polarization measured in 2016.
        The northern hemisphere is the best candidate region \citep{GRAY2026}.
        The resolved observations could also give constraints to the other possible theory that the higher polarization is caused by surface topography and observation geometry.
        DESTINY$^+$ observations will not be able to resolve local topography differences, such as those caused by particle sizes inferred in \citet{MACLENNAN2022}, but will be able to see if a certain region is more dominated by larger features, such as boulders or craters.
        The DESTINY$^+$ observations will also help to develop an accurate shadowing function for Phaethon's surface, which would enable the study of the effects of observation geometry.
        
        If our regolith model represents the true surface layer of Phaethon, the particles detected by DDA (DESTINY$^+$ Dust Analyzer) provide an independent test of particle populations inferred from the model.
        The model predicts a regolith consisting of large silicate-hosted particles mixed with much smaller iron-sulphide-rich aggregates, with a relative deficit of particles in the 1--100~{\textmu}m size range.
        Comparison with the particle sizes and compositions measured by DDA could test whether the dust environment of Phaethon is consistent with the surface layer inferred here from spectral and polarimetric observations.
        It will be especially interesting to see if the dust environment reflects the particle size range of our model, or if the dust environment reflects exactly 1--100~{\textmu}m particles, indicating preferential removal of particles in that size range from the surface.

\section{Conclusions}
\label{sec:conclusions}

    Our hypotheses for the composition of (3200) Phaethon's surface regolith based on light scattering modeling are the following:
    \begin{enumerate}
        \item The surface is composed of large ($>100$~{\textmu}m) opaque particles where silicate host includes darkening inclusions mixed with small ($<10$~{\textmu}m) darkening agent.
        \item The large silicate-hosted particle surfaces are highly fractured and full of small surface structure.
        \item The small darkening agent is organized into fluffy aggregate particles resembling interplanetary dust.
        \item The combination of large and small particles, and lack of particles in between, is the cause for the observed blue spectral slope.
    \end{enumerate}
    The modeled surface regolith resembles an anhydrous and thermally metamorphosed regolith composed of olivine and troilite.
    The closest meteorite analogue to the model regolith is an CM meteorite based on the troilite abundance in the model.
    
    The different spectral observations of Phaethon can be explained by relatively small changes in the troilite aggregate particle size distribution and do not require a change in composition. The higher polarization observed by \citet{ITO2018} can be explained by decreasing the contribution of the small troilite aggregate particles relative to the large olivine-hosted particles. The decrease in contribution may reflect either a true difference in the regolith or a local topography change in the effective contribution of the particle populations, which cannot be distinguished with the present model.
    
    DESTINY$^+$ will spatially resolve Phaethon's surface, but it will not resolve the regolith itself.
    DESTINY$^+$ will therefore face the same fundamental remote-sensing problems as ground-based observations that the observations do not uniquely reveal the particle sizes, composition, or regolith structure responsible for them. Rigorous light scattering modeling, that has been started here, will be essential for translating the new observations into physical regolith properties.
    
    The results from \cite{GRAY2026} predict intrinsic regolith heterogeneity between the northern and southern hemispheres.
    If the intrinsic hemispherical heterogeneity exists on Phaethon, the regolith model also predicts there to be a 20\% relative albedo difference between the hemispheres.
    DESTINY$^+$ will be able to detect such a difference, if it exists, and through it directly test our interpretations.

\begin{acknowledgments}
Research supported, in part, by the Research Council of Finland grants 336546, 359893, 361233, and 368493.
\end{acknowledgments}

\bibliography{sample701}{}
\bibliographystyle{aasjournalv7}

\end{document}